\documentclass[journal]{IEEEtran}
\usepackage{graphicx}
\usepackage{subfigure}
\usepackage{float}
\usepackage{amsmath}
\usepackage{epstopdf}
\usepackage{cases}
\usepackage{color}
\usepackage{cuted}\stripsep -3pt plus 3pt minus 2pt
\usepackage{algorithm}
\usepackage{algpseudocode}
\usepackage{amsmath}
\usepackage{amssymb}

\makeatletter
\renewcommand{\maketag@@@}[1]{\hbox{\m@th\normalsize\normalfont#1}}%
\makeatother
\usepackage{stfloats}
\usepackage{cite}
\usepackage{makecell}
\usepackage{multirow}
\usepackage{threeparttable}
\usepackage{booktabs}

\ifCLASSINFOpdf
\else
\fi

\usepackage{caption}
\usepackage{mathtools}

\begin{document}

\title{Iterative Signal Processing for Integrated Sensing and Communication Systems}

\author{Zhiqing Wei,
		Hanyang Qu,
		Wangjun Jiang, 
		Kaifeng Han, 
		Huici Wu,
		Zhiyong Feng
}

\maketitle

\begin{abstract}

Integrated sensing and communication (ISAC),
with sensing and communication sharing the same wireless resources and hardware,
has the advantages of high spectrum efficiency and low hardware cost,
which is regarded as one of the key technologies of the fifth generation advanced (5G-A) and
sixth generation (6G) mobile communication systems.
ISAC has the potential to be applied in the intelligent applications
requiring both communication and high accurate sensing capabilities.
The fundamental challenges of ISAC system are the
ISAC signal design and ISAC signal processing.
However, the existing ISAC signal has low anti-noise capability.
And the existing ISAC signal processing algorithms have the disadvantages of
quantization errors and high complexity, resulting in large energy consumption.
In this paper, phase coding is applied in ISAC signal design to improve
the anti-noise performance of ISAC signal.
Then, the effect of phase coding method on improving the sensing accuracy is analyzed.
In order to improve the sensing accuracy with low-complexity algorithm,
the iterative ISAC signal processing methods are proposed.
The proposed methods improve the sensing accuracy with low computational complexity,
realizing energy efficient ISAC signal processing.
Taking the scenarios of short distance and long distance sensing into account,
the iterative two-dimensional (2D) fast Fourier transform (FFT)
and iterative cyclic cross-correlation (CC) methods are proposed, respectively, 
realizing high sensing accuracy and low computational complexity.
Finally, the feasibility of the proposed ISAC signal processing methods
are verified by simulation results.
\end{abstract}


\IEEEpeerreviewmaketitle

\section{Introduction}

\subsection{Background}

With the rapid development of mobile communication systems
towards 5th-generation-advance (5G-A) and 6th-generation (6G),
new intelligent applications and services have emerged,
such as machine-type communication (MTC), 
smart city, and extended reality (XR) \cite{1},
which require both communication and sensing capabilities.
The separated design of sensing and communication will result in
the waste of scarce spectrum resources.
Integrated sensing and communication (ISAC),
with sensing and communication sharing the same wireless resources and hardware,
has the advantages of high spectrum efficiency and low hardware cost \cite{liu2020joint}.
Besides, the performance of sensing and communication will be improved via
the collaboration between sensing and communication \cite{weijie, liufanbeam}.
With ISAC applied in the intelligent applications in 5G-A and 6G,
high data rate communication and high accurate sensing could be achieved \cite{2}.
Actually, ITU-T's IMT-2030 has regarded ISAC as one of the key potential technologies
in the future mobile communication systems \cite{3}.

The fundamental challenges of ISAC system are the
ISAC signal design and ISAC signal processing.
There are significant differences in the signal structure
between communication signal and radar signal
due to different application scenarios.
The existing ISAC signals are classified into
communication centric design and radar centric design \cite{4}.
In the perspective of mobile communication systems,
the ISAC signal based on Orthogonal Frequency Division Multiplexing (OFDM)
attracted wide attention,
because OFDM is widely applied in mobile communication systems
and has excellent performance in both communication and sensing \cite{chiriyath2017radar}.
Since the transmit power of ISAC signal is relatively low compared with radar signal,
the power of echo signal of ISAC signal is low.
Hence, the anti-noise performance is crucial for ISAC signal design.
Besides, the ISAC signal processing with
high sensing accuracy and low computational complexity
is essential for ISAC system.
In this paper, the ISAC signal with anti-noise performance and the
ISAC signal processing with high sensing accuracy and low computational complexity
are studied.

\subsection{Related Works}

In order to improve the anti-noise capability of ISAC signal,
the ISAC signal design schemes based on
spread spectrum, linear frequency modulation (LFM), and phase coding, are proposed.
Chen \emph{et al.}
combined direct spread spectrum technology and OFDM,
designing the ISAC signal with anti-noise performance \cite{chenxv}.
The ISAC signal combining OFDM and LFM obtains large time-bandwidth product,
improving the anti-noise performance of
ISAC signal.
\cite{chen2015study} applied the orthogonal
LFM signal as the subcarrier and adopts fractional Fourier transform (FRFT) as
the modulation and demodulation method.
This scheme effectively reduces bit error rate (BER) in Rayleigh channel.
However, the performance of sensing is not improved.
Phase coding was applied to ISAC signal design in \cite{uysal2019phase},
which improves the correlation of ISAC signals,
thereby improving the anti-noise performance of ISAC signals.
The ISAC signal combining OFDM and spread spectrum requires large bandwidth, 
which is difficult for mobile communication systems because of the scarce spectrum resources. 
The ISAC signal combining OFDM and LFM 
is not compatible to the current mobile communication systems. 
The ISAC signal combining OFDM and phase coding effectively improves the anti-noise performance
and is compatible to the modern mobile communication systems.

ISAC signals based on phase coding have been studied extensively. 
The main phase coded sequences include pseudo-random 
sequences \cite{hu2014radar,8,tian2017radar}, 
chaotic sequences \cite{zhao2014chaos}, 
complete complementary codes \cite{9} and Golay sequences \cite{6,7}.
Hu \emph{et al.} applied random sequences to 
design ISAC signals satisfying the requirements of 
sensing and communication \cite{hu2014radar}. 
The deserialized digital sequence is modulated 
in the corresponding subcarriers, 
and the information required for sensing and communication is extracted at the RX. 
A phase-coded OFDM signal design for a single scattering point target was proposed in \cite{8}. 
Then, they proposed a high-resolution signal processing algorithm \cite{tian2017radar}. 
However, the pseudo-random sequences such as m-sequences and Gold sequences do not have ideal cross-correlation. 
Moreover, there are a limited number of sequence types.
Therefore, some studies have paid much attention on chaotic sequences. 
Zhao \emph{et al.} designed a phase-coding OFDM signal using chaotic sequences, 
which has a large time-bandwidth product \cite{zhao2014chaos}. 
They further extracted phase code sequence to improve correlation. 
Besides, the ISAC signal design using chaotic sequences 
has high flexibility to support different scenarios \cite{zhao2014chaos}. 
However, 
chaotic sequences have satisfactory correlation only when there are enough coded bits.
Recently, Qi \emph{et al.} proposed the phase coding scheme with complete complementary codes, 
achieving high data rate and accurate target sensing \cite{9}.
\cite{6} compared the phase coded sequences and discoverd that
Golay sequences restrict the peak to average power ratio (PAPR) of ISAC signal 
with the increase of the number of subcarriers.
\cite{7} applied Reed Muller codes to encode random communication bits into Golay sequences, 
which not only reduces the PAPR of ISAC signal, 
but also improves the error detection and correction capability of communication. 
Considering the coding efficiency, 
PAPR, bit error rate, etc., 
the Golay sequences are chosen as the coding sequences in this paper.

Since the structure of ISAC signal is different from traditional radar signal,
it is necessary to design the signal processing method suitable
to ISAC signal.
Sturm \textit{et al.}
proposed the two-dimensional fast Fourier transform (2D FFT) method for 
ISAC signal processing \cite{12}.
However, there is quantization error in 2D FFT method.
Besides, the maximum detection distance of OFDM signal is limited by
the length of cyclic prefix (CP).
In order to solve this problem,
Wu \textit{et al.} in \cite{14} applies virtual cyclic prefix (VCP) 
into 2D FFT method to estimate the distance and velocity of target.
The signal processing method proposed in \cite{13}
improves the anti-noise performance and
sensing accuracy of ISAC signal by increasing the number of samplings.
However, the quantization error in 2D FFT method is relatively large in the 
above methods, which will degrade the performance of ISAC signal processing.

The resolution of distance and velocity estimation in 2D FFT method is limited by 
the number of subcarriers and the number of OFDM symbols.
In order to reduce the quantization error, 
Li \textit{et al.} applied fractional Fourier transform (FRFT) to 2D FFT method,
proposing 2D FRFT method.
The 2D FRFT method increases the number of the points of Fourier transform, 
so that the quantization error is reduced and the sensing accuracy is improved \cite{15}.
However, the computational complexity in \cite{15} is high.
\cite{16} designed 2D FFT method for short distance sensing
and correlation method with Golay sequences in the preamble for long distance sensing.
The proposed scheme realizes high accuracy of distance and velocity estimation.
Since the duration of the preamble is very short,
multiple frames are required to achieve high accuracy of velocity estimation \cite{16}.
The super resolution methods such as 
multiple signal classification (MUSIC) \cite{schmidt1986multiple} and 
estimating signal parameter via rotational invariance technique (ESPRIT) \cite{roy1986esprit,tureli2000ofdm} methods 
bring large computational complexity to ISAC systems.
Overall, 2D FFT method faces the 
challenge of large quantization error. 
FRFT and super resolution methods face the challenge of large overhead and 
computational complexity. 
Therefore, the ISAC signal processing method with low complexity and high accuracy is required.

\subsection{Our Contributions}

\begin{figure*}[!htbp]
	\centering
	\includegraphics[width=0.98\textwidth]{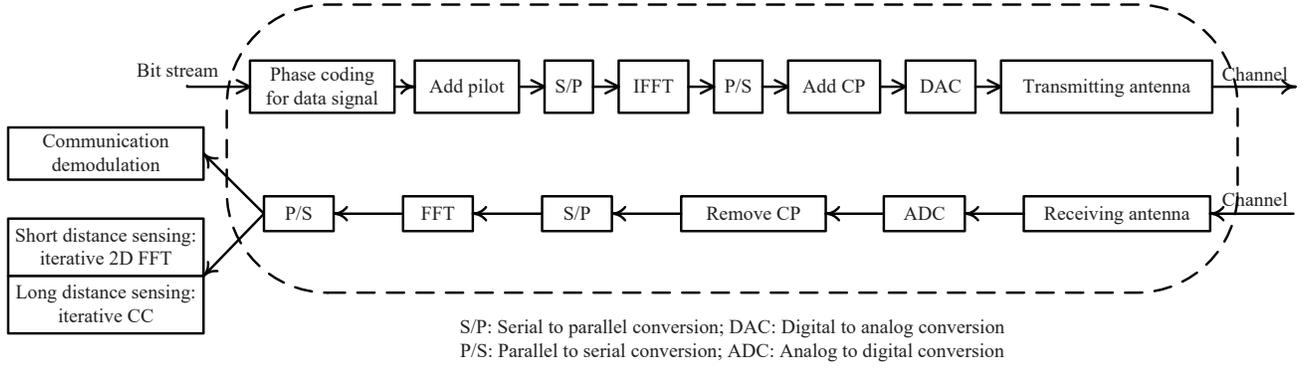}
	\caption{System model of PC-OFDM ISAC signal.}
	\label{system}
\end{figure*}

In order to design the ISAC signal with anti-noise capability,
considering pilot and data symbols in a typical 5G New Radio (NR) frame,
the phase coding is applied in the data symbols.
Then, the pilot and phase coded data symbols are jointly applied in target sensing.
The iterative ISAC signal processing methods with 
high accuracy and low complexity
are proposed.
The sensing distance is limited by the length of CP,
so that the scenarios of short distance sensing and long distance sensing are considered. 
In the scenario of short distance sensing, 
the delay of the echo signal is smaller than the 
length of CP. 
While in the scenario of long distance sensing, 
the delay of the echo signal is larger 
than the 
length of CP. 
The iterative 2D FFT and iterative Cyclic Cross-correlation (CC) 
methods are designed for short distance sensing and long distance sensing, respectively.
The main contributions of this paper are summarized as follows.

\begin{itemize}
	\item The Phase Coding-OFDM (PC-OFDM) 
	ISAC signal with the pilot and phase coded data symbols is designed.
	In long range sensing, phase encoding improves signal correlation and 
	signal to noise ratio (SNR),
	enhancing the performance of sensing.
	Then, we analyze the CRLB of distance and velocity estimation.
	\item The iterative 2D FFT and iterative 
	CC methods are designed for the scenarios of short distance sensing and 
	long distance sensing, respectively, 
	realizing the sensing 
	with low complexity and high accuracy. 
	The estimated value obtained in each iteration compensates the phase 
	of the channel information matrix. 
	With the increase of the number of iterations, 
	the sensing accuracy is improving.
\end{itemize}

\subsection{Outline of This Paper}

The remaining parts of this paper are organized as follows.
Section \uppercase\expandafter{\romannumeral2} proposes the  
PC-OFDM ISAC signal.
Section \uppercase\expandafter{\romannumeral3} 
designs the iterative 2D FFT method for short distance sensing.
In Section \uppercase\expandafter{\romannumeral4}, the iterative CC method
is designed for long distance sensing. 
The simulation results are shown in Section \uppercase\expandafter{\romannumeral5}. 
Section \uppercase\expandafter{\romannumeral6} summarizes this paper.

\section{ISAC Signal Model}

The block diagram of the ISAC signal is shown in Fig. \ref{system}.
At the transmitter (TX),
a phase coding module is added to encode the bit stream.
Then the phase coded data signal is combined with the pilot signal 
for target sensing.
At the receiver (RX), the echo signal is 
sent to the communication demodulation module to obtain
the communication
information and to the target sensing module
to obtain the sensing information
of target.
In the scenario of short distance sensing, the 
iterative 2D FFT method is designed.
While in the scenario of long distance sensing, 
the iterative CC method is proposed.

5G NR specifies that the frame of downlink signal includes
pilot signal and data signal.
In ISAC system, the base station (BS) 
obtains the distance and velocity of the target
by processing the echo of the downlink transmitted signal.
The pilot signal is written as follows.

\begin{equation}\label{eq1}
{s_p}\left( t \right) = \sum\limits_{m = 0}^{{M_p} - 1} {\sum\limits_{n = 0}^{{N_p} - 1} {{a_{nm}}{e^{j2\pi n\Delta {f_p}t}}{\mathop{\rm rect}\nolimits} \left( {\frac{{t - m{T_{sym}}}}{{{T_{sym}}}}} \right)} },
\end{equation}
where ${M_p}$ is the number of symbols occupied by the pilot signal,
${N_p}$ is the number of subcarriers occupied by the pilot signal,
${a_{nm}}$ is the modulation symbol of the pilot signal modulated on the $n$-th subcarrier of the $m$-th OFDM symbol,
$\Delta {f_p}$ is the interval of subcarrier of the pilot signal,
${T_{sym}}$ is the duration of OFDM symbol.

The data signal is

\begin{equation}\label{eq2}
{s_d}\left( t \right) = \sum\limits_{m = 0}^{{M_d} - 1} {\sum\limits_{n = 0}^{{N_d} - 1} {{d_{nm}}{e^{j2\pi n\Delta ft}}{\mathop{\rm rect}\nolimits} \left( {\frac{{t - m{T_{sym}}}}{{{T_{sym}}}}} \right)} },
\end{equation}
where ${M_d}$ is the number of symbols occupied by the data signal,
${N_d}$ is the number of subcarriers occupied by the data signal,
${d_{nm}}$ is the modulation symbol of the data signal modulated on the $n$-th subcarrier of the $m$-th OFDM symbol,
$\Delta {f}$ is the interval of subcarrier of the data signal.
Generally, the interval of subcarrier of the pilot signal is the same as that of data signal,
namely $\Delta {f_p} = \Delta {f}$.

With the combination of pilot and data signals, 
according to (\ref{eq1}) and (\ref{eq2}),
the transmitted signal is

\begin{equation}
s\left( t \right) = {s_p}\left( t \right) + {s_d}\left( t \right).
\end{equation}

\begin{figure*}[!ht]
	\normalsize
	\begin{equation}\label{eq12}
		\begin{array}{l}
			- E\left( {\frac{{{\partial ^2}\bar L\left( {\theta \left| r \right.} \right)}}{{\partial {\varepsilon ^2}}}} \right) = \frac{L}{{{\sigma ^2}}}\sum\limits_{\tilde m = 0}^{\tilde M - 1} {} \left\{ {\frac{1}{3}\sum\limits_{i = 0}^{\tilde Q - 1} {\left( {\frac{{\partial {s_{\tilde m}}\left[ i \right]}}{{\partial \varepsilon }} \cdot \frac{{\partial s_{_{\tilde m}}^*\left[ i \right]}}{{\partial \varepsilon }}} \right)} } \right. \left. { + \frac{1}{2}\sum\limits_{i = \tilde Q}^{\tilde N - 1} {\left( {\frac{{\partial {s_{\tilde m}}\left[ i \right]}}{{\partial \varepsilon }} \cdot \frac{{\partial s_{_{\tilde m}}^*\left[ i \right]}}{{\partial \varepsilon }}} \right)} } \right\}
		\end{array},
	\end{equation}
	
	\begin{equation}\label{eq13}
		\begin{array}{l}
			- E\left( {\frac{{{\partial ^2} \bar L\left( {\theta \left| r \right.} \right)}}{{\partial {\upsilon ^2}}}} \right) = \frac{L}{{{\sigma ^2}}}\sum\limits_{\tilde m = 0}^{\tilde M - 1} {} \left\{ {\frac{1}{3}\sum\limits_{i = 0}^{\tilde Q - 1} {\left( {\frac{{\partial {s_{\tilde m}}\left[ i \right]}}{{\partial \upsilon }} \cdot \frac{{\partial s_{_{\tilde m}}^*\left[ i \right]}}{{\partial \upsilon }}} \right)} } \right. \left. { + \frac{1}{2}\sum\limits_{i = \tilde Q}^{\tilde N - 1} {\left( {\frac{{\partial {s_{\tilde m}}\left[ i \right]}}{{\partial \upsilon }} \cdot \frac{{\partial s_{_{\tilde m}}^*\left[ i \right]}}{{\partial \upsilon }}} \right)} } \right\}
		\end{array},
	\end{equation}
	
	\begin{equation}\label{eq14}
		- E\left( {\frac{{{\partial ^2}\bar L\left( {\theta \left| r \right.} \right)}}{{\partial \varepsilon \partial \upsilon }}} \right) = \frac{L}{{4{\sigma ^2}}}\sum\limits_{\tilde m = 0}^{\tilde M - 1} {\left[ \begin{array}{l}
				\frac{2}{3}\sum\limits_{i = 0}^{\tilde Q - 1} {\left( {\frac{{\partial {s_{\tilde m}}\left[ i \right]}}{{\partial \varepsilon }}\frac{{\partial s_{\tilde m}^*\left[ i \right]}}{{\partial \upsilon }} + \frac{{\partial s_{\tilde m}^*\left[ i \right]}}{{\partial \varepsilon }}\frac{{\partial {s_{\tilde m}}\left[ i \right]}}{{\partial \upsilon }}} \right)} \\
				+ \sum\limits_{i = \tilde Q}^{\tilde N - 1} {\left( {\frac{{\partial {s_{\tilde m}}\left[ i \right]}}{{\partial \varepsilon }}\frac{{\partial s_{\tilde m}^*\left[ i \right]}}{{\partial \upsilon }} + \frac{{\partial s_{\tilde m}^*\left[ i \right]}}{{\partial \varepsilon }}\frac{{\partial {s_{\tilde m}}\left[ i \right]}}{{\partial \upsilon }}} \right)}
			\end{array} \right]}.
	\end{equation}
\end{figure*}

Then, the ISAC signal is obtained by combining the pilot signal and the phase coded data signal.
The modulation symbol after phase coding is denoted by ${d'_{nm}}$.
Hence, the phase coded data signal in the ISAC signal is

\begin{equation}
	\begin{aligned}
		{s_d}\left( t \right) = \sum\limits_{m = 0}^{{M_d} - 1} {\sum\limits_{n = 0}^{{N_d} - 1} {{{d'}_{nm}}{e^{j2\pi n\Delta ft}}} } {\mathop{\rm rect}\nolimits} \left( {\frac{{t - m{T_{sym}}}}{{{T_{sym}}}}} \right).
	\end{aligned}
\end{equation}

The transmitted signal is reflected by the target.
The echo signal is received by the RX.
Compared with the transmitted signal,
the echo signal includes delay $\tau$ and Doppler frequency shift ${f_d}$,
with expression as follows.

\begin{equation}
	\begin{aligned}
		&	r( t ) = hs( {t - \tau } ){e^{j2\pi {f_d}t}} = h \times \\
		&	 \sum\limits_{m = 0}^{M - 1} {\sum\limits_{n = 0}^{N - 1} {d( {n,m} ){e^{j2\pi n\Delta f( {t - \tau } )}}{e^{j2\pi {f_d}t}}{\mathop{\rm rect}\nolimits} ( {\frac{{t - \tau  - m{T_{sym}}}}{T_{sym}}} )} } \\
		& +\omega \left( t \right),
	\end{aligned}
\end{equation}
where $h$ represents the product of channel attenuation 
and coefficient of radar cross-section (RCS), 
$\omega \left( t \right)$ is Additive White Gaussian Noise (AWGN),
$M = M_p + M_d$ is the number of symbols, 
$N$ is the number of subcarriers, $d(n,m)$ is the modulation symbol on the $n$-th subcarrier of the $m$-th OFDM symbol in the ISAC signal combining the pilot signal and the phase coded data signal.

In long distance sensing, phase coding is 
applied to enhance the anti-noise performance of ISAC signal.
The transmitted ISAC signal $s\left( t \right)$ and 
received ISAC signal $r\left( t \right)$ are sampled.
The sampling points of them are divided into $\tilde M$ groups.
The number of sampling points in each group is $\tilde N$ \cite{14}.
The $i$-th sampling point of the $\tilde m$-th group of $s\left( t \right)$ 
is denoted by ${s_{\tilde m}}\left[ i \right]$.
The $i$-th sampling point of the $\tilde m$-th group of $r\left( t \right)$ is 
denoted by ${r_{\tilde m}}\left[ i \right]$.
The VCP with length $\tilde Q$ is added \cite{14}.
Then, the performance improvement of phase coding on the ISAC signal 
is demonstrated by deriving the CRLB.
The estimation vector $\theta$ composed of $\varepsilon  = \frac{\tau }{{{T_b}}}$ and $\upsilon  = {f_d}{T_b}$ is written as follows.

\begin{equation}
	\theta  = {\left[ {\varepsilon ,\upsilon } \right]^T}.
\end{equation}

The Fisher information matrix is
\begin{equation}
	{\bf{J}}\left( \theta  \right) = \left[ {\begin{array}{*{20}{c}}
			{ - E\left[ {\frac{{{\partial ^2}\bar L\left( {\theta \left| r \right.} \right)}}{{\partial {\varepsilon ^2}}}} \right]}&{ - E\left[ {\frac{{\partial ^2 {{\bar L}}\left( {\theta \left| r \right.} \right)}}{{\partial \varepsilon \partial \upsilon }}} \right]}\\
			{ - E\left[ {\frac{{\partial ^2 {{\bar L}}\left( {\theta \left| r \right.} \right)}}{{\partial \upsilon \partial \varepsilon }}} \right]}&{ - E\left[ {\frac{{\partial ^2 {{\bar L}}\left( {\theta \left| r \right.} \right)}}{{\partial {\upsilon ^2}}}} \right]}
	\end{array}} \right],
\end{equation}
where $\bar L\left( { \theta  \left|r \right.} \right)$ is the log-likelihood function,
$r$ is the echo signal. 
The log-likelihood function is constructed according to the samplings of 
the received echo signal,
which is shown as follows.
\begin{equation}
	\begin{aligned}
		& \bar L\left( {\left. \theta  \right|r} \right) = \ln \left( {\prod\limits_{\tilde m = 0}^{\tilde M - 1} ( \prod\limits_{i = 0}^{\tilde Q - 1} {{{\left( {\frac{1}{{2\pi  \times 3\frac{{{\sigma ^2}}}{L}}}} \right)}^{\frac{1}{2}}}{e^{ - \frac{{{{\left( {{r_{\tilde m}}\left[ i \right] - {{\tilde s}_{\tilde m}}} \right)}^2}}}{{2 \times 3\frac{{{\sigma ^2}}}{L}}}}}} } \right.\\
		& \left. { \cdot \prod\limits_{i = \tilde Q}^{\tilde N - 1} {{{\left( {\frac{1}{{2\pi  \times 2\frac{{{\sigma ^2}}}{L}}}} \right)}^{\frac{1}{2}}}{e^{ - \frac{{{{\left( {{r_{\tilde m}}\left[ i \right] - {{\tilde s}_{\tilde m}}} \right)}^2}}}{{2 \times 2\frac{{{\sigma ^2}}}{L}}}}}} } \right)\\
		= & \frac{{\tilde Q\tilde M}}{2}\ln \frac{2}{3} + \frac{{\tilde N\tilde M}}{2}\ln \frac{L}{{4\pi {\sigma ^2}}}\\
		& - \frac{L}{{{\rm{6}}{\sigma ^2}}}\sum\limits_{\tilde m = 0}^{\tilde M - 1} {\left( {\sum\limits_{i = 0}^{\tilde Q - 1} {\left( {{r_{\tilde m}}\left[ i \right] - {{\tilde s}_{\tilde m}}\left[ i \right]} \right)\left( {r_{\tilde m}^*\left[ i \right] - \tilde s_{\tilde m}^*\left[ i \right]} \right)} } \right.}  \\
		& + \left. {\sum\limits_{i = \tilde Q}^{\tilde N - 1} {\left( {{r_{\tilde m}}\left[ i \right] - {{\tilde s}_{\tilde m}}\left[ i \right]} \right)\left( {r_{\tilde m}^*\left[ i \right] - \tilde s_{\tilde m}^*\left[ i \right]} \right)} } \right),
	\end{aligned}
\end{equation}
\begin{equation}
	{{\tilde s}_{\tilde m}}\left[ i \right] = h{s_{\tilde m}}\left[ {i - \varepsilon } \right]{e^{j2\pi \upsilon \left( {\tilde m\tilde N + i} \right)}},
\end{equation}
\begin{equation}
	L = 10\log \left( {\tilde N} \right),
\end{equation}
where $\sigma ^2$ is the power of AWGN,
$L$ is the SNR gain obtained by coherent demodulation of the phase coded signal at RX, which is related to the length of the phase code participating in coherent accumulation.

The first derivative and 
the second derivative of the estimation vector $\theta$ are obtained by the log-likelihood function.
The elements in Fisher information matrix are given as (\ref{eq12}) (\ref{eq13}) (\ref{eq14})
at the top of this page, 
where $s_{\tilde m}^ * \left[ i \right]$ is the conjugate of ${s_{\tilde m}}\left[ i \right]$.
CRLB is the inverse of Fisher information matrix. 
The CRLB of $\varepsilon $ and $\upsilon $ are as follows.

\begin{equation}
	\begin{aligned}
		{I_{CRLB,\varepsilon}} & = \frac{{{\bf{J}}\left( {2,2} \right)}}{{{\bf{J}}\left( {1,1} \right){\bf{J}}\left( {2,2} \right) - {\bf{J}}\left( {1,2} \right){\bf{J}}\left( {2,1} \right)}} \\
		& = \frac{6}{{{\pi ^2}{{\left| h \right|}^2}L \cdot SNR\left( {\tilde N - \frac{{\tilde Q}}{3}} \right)N\left( {4 - {{\left( {\frac{{\tilde N}}{N}} \right)}^2}} \right)}},
	\end{aligned}
\end{equation}
\begin{equation}
	\begin{aligned}
		{I_{CRLB,\upsilon}} & = \frac{{{\bf{J}}\left( {1,1} \right)}}{{{\bf{J}}\left( {1,1} \right){\bf{J}}\left( {2,2} \right) - {\bf{J}}\left( {1,2} \right){\bf{J}}\left( {2,1} \right)}} \\
		& = \frac{4}{{3{\pi ^2}{{\left| h \right|}^2}L \cdot SNR\left( {\tilde N - \frac{{\tilde Q}}{3}} \right)N\tilde M\left( {\tilde M - 1} \right)}},
	\end{aligned}
\end{equation}
where ${{\bf{J}}\left( {m,n} \right)}$ is the element of $m$-th row and $n$-th column in 
matrix ${\bf{J}}$, $SNR$ is the SNR of the received echo signal.
According to the relation between distance and $\varepsilon$, 
and the relation between velocity and $\upsilon$, 
the CRLBs of distance and velocity estimation are derived as follows.

\begin{equation}
	\begin{array}{l}
		{I_{CRLB,R}} = \frac{{3T_b^2{c^2}}}{{2{\pi ^2}{{\left| h \right|}^2}L \cdot SNR}}\cdot \frac{1}{{\left( {\tilde N - \frac{{\tilde Q}}{3}} \right)N\left( {4 - {{\left( {\frac{{\tilde N}}{N}} \right)}^2}} \right)}}
	\end{array},
\end{equation}
\begin{equation}
	\begin{array}{l}
		{I_{CRLB,V}} = \frac{{{c^2}}}{{3{\pi ^2}{{\left| h \right|}^2}L \cdot SNR \cdot T_b^2f_c^2}}\cdot \frac{1}{{\left( {\tilde N - \frac{{\tilde Q}}{3}} \right)N\tilde M\left( {\tilde M - 1} \right)}},
	\end{array}.
\end{equation}
where $T_b$ is sampling interval,
$c$ is the propagation velocity of light, 
$f_c$ is carrier frequency.
It is revealed that CRLB is related to the SNR gain $L$ due to coherent accumulation benefited from phase coding.
Hence, the phase coding effectively
improves the anti-noise performance of ISAC signal.

\section{Short Distance Sensing}

This section studies the ISAC signal processing algorithm
in the scenario of short distance sensing.
The pilot and data signals are both applied for target sensing.
The 2D FFT method is applied to process the ISAC signal.
Besides, the iterative 2D FFT algorithm is designed to achieve
the ISAC signal processing with high accuracy and low complexity,
which effectively reduces the quantization error caused by 2D FFT method.

\subsection{2D FFT Method}
\label{sec3a}

When receiving the echo signal,
the RX applies 2D FFT method to process the echo signal.
2D FFT method was firstly proposed by Strum \emph{et al.}, 
which has the advantage of low computational complexity. The 2D FFT method proposed in \cite{12} is applied for ISAC signal processing. Since the pilot signal has good correlation and stronger power compared with the data signal, the pilot signal is selected for the first delay estimation. Besides, the pilot signal and data signal are combined for the velocity estimation. 
Firstly, the echo signal is sampled with the sampling interval ${T_{sym}}$ 
in time domain and the sampling interval $\Delta f$ in frequency domain to
obtain the sampling result ${d_{Rx}}\left( {mM + n} \right)$. 
According to the samplings of transmitted signal
${d_{Tx}}\left( {mM + n} \right)$ and
received signal ${d_{Rx}}\left( {mM + n} \right)$,
the transmitted symbol matrix ${{\bf{T}}_r}$ and
received symbol matrix ${{\bf{R}}_e}$ are constructed.
The channel information matrix ${\bf{Y}}$ is obtained by
dividing the received symbol matrix ${{\bf{R}}_e}$ by
the transmitted symbol matrix ${{\bf{T}}_r}$ element by element.
We have ${\bf{Y}} = {{\bf{k}}_r} \otimes {{\bf{k}}_v}$ with $\otimes$
representing the Kronecker product.
${\textbf{k}_r}$ and ${\textbf{k}_v}$ are vectors 
about time delay and Doppler freuency shift respectively,
which are written as follows.

\begin{equation}
	\begin{aligned}
		{{\bf{k}}_r} = [1,{e^{ - j2\pi \Delta f\tau }}, \cdots ,{e^{ - j2\pi \left( {{N_p} - 1} \right)\Delta f\tau }}],
	\end{aligned}
\end{equation}

\begin{equation}
	\begin{aligned}
		{{\bf{k}}_v} = [1,{e^{j2\pi {T_{sym}}{f_d}}}, \cdots ,{e^{j2\pi \left( {{M_p} + {M_d} - 1} \right){T_{sym}}{f_d}}}].
	\end{aligned}
\end{equation}

In the next subsection, the ISAC signal combining pilot and 
phase coded data signal is applied, which 
improves the sensing performance because of a large number of OFDM symbols and subcarriers.

The peak indexes of the 2D FFT of ${\bf{Y}}$, denoted by $k_1$ and ${l_1}$, are obtained,
revealing the delay and Doppler freuency shift,
so as to obtain the initial estimation of distance and velocity of the target as follows \cite{12}.

\begin{equation}
	R = \frac{c}{{2{N_p}\Delta {f_p}}}{k_1} = \frac{c}{{2B}}{k_1},
\end{equation}
\begin{equation}
	V = \frac{{c{l_1}}}{{2{f_c}{T_{sym}}\left( {{M_p} + {M_d}} \right)}},
\end{equation}
where $B={{N_p}\Delta {f_p}}$ is the bandwidth of ISAC signal.
There are still a large quantization error in 2D FFT method.
Thus, the iterative 2D FFT method is designed to improve
the sensing accuracy with low complexity.

\subsection{Iterative 2D FFT Method}
\label{sec3b}

\begin{figure}
	\centering
  \includegraphics[width=0.36\textwidth]{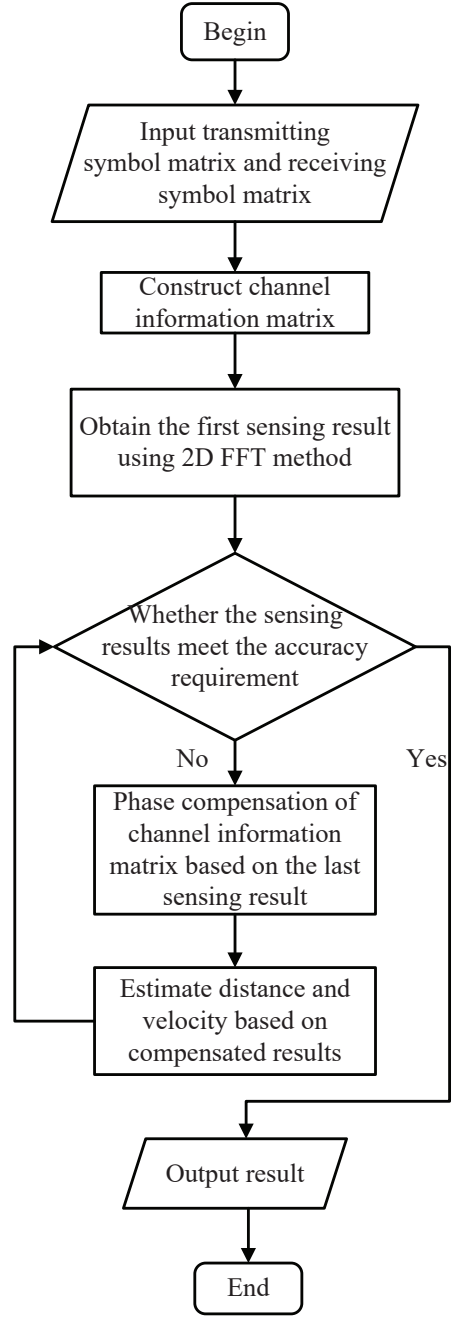}
  \caption{Iterative 2D FFT method.}
  \label{shortdistancehighaccuracy}
\end{figure}

This section proposes iterative 2D FFT method for
the scenario of short distance sensing,
reducing the quantization error caused by 2D FFT method.
The algorithm is shown in Fig. \ref{shortdistancehighaccuracy}.
The channel information matrix is obtained via dividing the transmitted symbol matrix 
by the received symbol matrix element by element. 
Then, the first estimation result is obtained by 2D FFT. 
If the sensing accuracy does not meet the requirement, 
the phase of the channel information matrix is compensated based on the first estimation result. 
The second estimation result is obtained based on phase compensation. 
The above processes are executed iteratively 
until the sensing accuracy satisfies the requirement.

Take distance estimation as an example. The delay $\tau$ is written as

\begin{equation}
	\tau = \sum\limits_{i = 1}^X {{\tau _i}},
\end{equation}
where $\tau _i$ is the estimated delay in $i$-th iteration,
$X$ is the number of iterations.
With the initial delay estimation,
the $\tau_i$ is as follows.

\begin{equation}\label{eq23}
	{\tau _i} = \left\{ {\begin{array}{*{20}{l}}
			{\frac{{{k_i} - 0.5}}{{\Delta {f_p}{N_p}N_d^{i - 1}}},i < X}\\
			{\frac{{{k_i}}}{{\Delta {f_p}{N_p}N_d^{i - 1}}},i = X}
	\end{array}} \right..
\end{equation}

In (\ref{eq23}), 
$k_i$ is the peak index obtained in $i$-th iteration. 
In Section \ref{sec3a}, the peak index $k_1$ is obtained. 
Thus $\tau_1$ is derived via (\ref{eq23}).
The $m$-th column vector in channel information matrix ${\bf{Y}}$ is written as (\ref{eq24}). 
The phase compensation of the delay term of $m$-th column vector in ${\bf{Y}}$ 
is carried out according to the ${\tau _1}$ 
obtained in the first iteration,
yielding ${{\textbf{k}^\prime_{rm}} }$ as shown in (\ref{eq25}).
The IDFT of ${{\textbf{k}^\prime_{rm}} }$ is calculated.
By searching the peak of IDFT of $\textbf{k}^\prime_{rm}$,
the peak index is obtained to estimate the distance of target.
The column vector ${{\bf{k}}_{rm}}$ comes from the data signal, 
since the there are more subcarriers in data signal compared with the pilot signal. 

\begin{equation}\label{eq24}
{{\bf{k}}_{rm}} = {\left[ {1,{e^{ - j2\pi \Delta f\tau }}, \cdots ,{e^{ - j2\pi \left( {{N_d} - 1} \right)\Delta f\tau }}} \right]^T}.
\end{equation}

\begin{equation}\label{eq25}
{\bf{k}}_{rm}^\prime  = {{\bf{k}}_{rm}} \odot {\left[ {1,{e^{j2\pi \Delta f{\tau _1}}}, \cdots ,{e^{j2\pi \left( {{N_d} - 1} \right)\Delta f{\tau _1}}}} \right]^T},
\end{equation}
where $\odot$ is the hadamard product.

The IDFT of ${{\textbf{k}^\prime_{rm}} }$ is calculated as follows.
\begin{equation}\label{eq26}
	\begin{aligned}
		{\bf{r}}\left( {{{\bar k}_d}} \right) & = \frac{1}{{{N_d}}}\sum\limits_{n = 0}^{{N_d} - 1} {{e^{ - j2\pi n\Delta f\left( {\tau  - {\tau _1}} \right)}}{e^{j2\pi n\frac{{{\bar k}_d}}{{{N_p}{N_d}}}}}}\\
		& = \frac{1}{{{N_d}}}\sum\limits_{n = 0}^{{N_d} - 1} {{e^{ - j2\pi n\Delta f{\tau _2}}}{e^{j2\pi n\frac{{{\bar k}_d}}{{{N_p}{N_d}}}}}} \\
		& = \frac{1}{{{N_d}}}\sum\limits_{n = 0}^{{N_d} - 1} {{e^{j2\pi n\left( {\frac{{{\bar k}_d}}{{{N_p}{N_d}}} - \Delta f{\tau _2}} \right)}}},
	\end{aligned}
\end{equation}
where ${{{\bar k}_d}} = 0,1,...,{N_d} - 1$,
${\bf{r}}\left( {{{\bar k}_d}} \right)$ is the ${\bar k}_d$-th element 
of the IDFT of ${{\textbf{k}^\prime_{rm}} }$. 
When there are only 
two iterations, we have ${\tau  - {\tau _1}}={\tau _2}$.

Denoting $\Delta \tau  = \frac{{{\bar k}_d}}{{{N_p}{N_d}}} - \Delta f{\tau _2}$, 
(\ref{eq26}) is transformed as follows.

\begin{equation}\label{eq27}
	\begin{aligned}
		{\bf{r}}\left( {\Delta \tau } \right) & = \frac{1}{{{N_d}}}\sum\limits_{n = 0}^{{N_d} - 1} {{e^{j2\pi n\Delta \tau }}} \\
		& = \frac{1}{{{N_d}}}\frac{{1 - {e^{j2\pi \Delta \tau {N_d}}}}}{{1 - {e^{j2\pi \Delta \tau }}}}\\
		& = \frac{1}{{{N_d}}}\frac{{{e^{ - j\pi \Delta \tau {N_d}}} - {e^{j\pi \Delta \tau {N_d}}}}}{{\left( {{e^{ - j\pi \Delta \tau }} - {e^{j2\pi \Delta \tau }}} \right){e^{ - j\pi \Delta \tau \left( {{N_d} - 1} \right)}}}}\\
		& = \frac{1}{{{N_d}}}\frac{{\sin \left( {\pi \Delta \tau {N_d}} \right)}}{{\sin \left( {\pi \Delta \tau } \right)}}{e^{j\pi \Delta \tau \left( {{N_d} - 1} \right)}}\\
		& \approx {\rm{sinc}}(\pi \Delta \tau {N_d}){e^{j\pi \Delta \tau \left( {{N_d} - 1} \right)}}.
	\end{aligned}
\end{equation}

In terms of ${\mathop{\rm sinc}\nolimits} \left( {\pi \Delta \tau {N_d}} \right)$
and $\exp \left( {j\pi \Delta \tau \left( {{N_d} - 1} \right)} \right)$ in (\ref{eq27}),
when $\Delta \tau = 0$, both of them are maximum.
Therefore, when $\Delta \tau = 0$, 
${\bf{r}}\left( {{{\bar k}_d}} \right)$ is maximum. 
Searching the peak of ${\bf{r}}\left( {{{\bar k}_d}} \right)$, 
the peak index $k_2$ is obtained, which satisfies the following equation.

\begin{equation}\label{eq28}
	\Delta \tau = \frac{{k_2}}{{{N_p}{N_d}}} - \Delta f{\tau _2} = 0.
\end{equation}
Solving (\ref{eq28}), ${{\tau _2} = \frac{{{k_2}}}{{\Delta f{N_p}{N_d}}}}$ is obtained. 
With two iterations, 
the distance of the target is derived as follows.
\begin{equation}
{R = \frac{c}{2}\left( {{\tau _1} + {\tau _2}} \right) = \frac{c}{{2\Delta f}}\frac{{{k_1} - \frac{1}{2} + \frac{{{k_2}}}{{{N_p}}}}}{{{N_d}}}}	
\end{equation}

\begin{figure}
  \centering
  \includegraphics[width=0.49\textwidth]{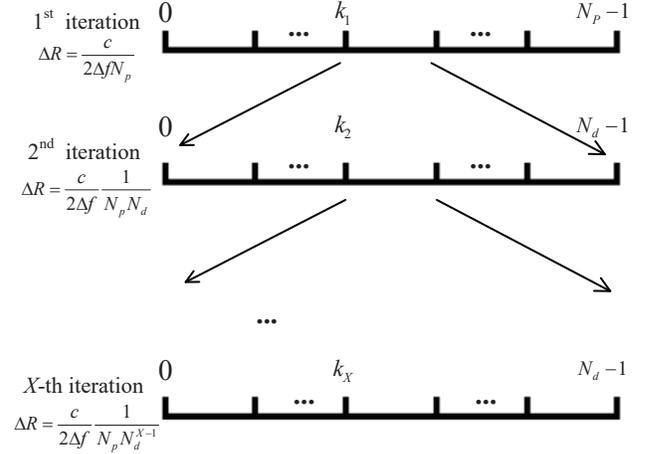}
  \caption{The iterations of distance estimation.}
  \label{Riterate}
\end{figure}

With $X$ iterations, 
the accuracy of distance estimation is enhanced. 
The iteration process is shown in the Fig. \ref{Riterate}.
Compared with the distance estimation of 2D FFT method,
the accuracy of distance estimation with $X$ iterations is

\begin{equation}
	\Delta R = \frac{c}{{2{\Delta f}}} \cdot \frac{1}{{{N_p}N_d^{X - 1}}}.
\end{equation}

Similarly, 
the above iterative 2D FFT method is applied in velocity estimation.
With two iterations, Doppler frequency shift is ${f_d} = {f_{d1}}+{f_{d2}}$,
where $f_{di}$ is the estimated Doppler frequency shift in the $i$-th iteration.
When estimating velocity in the first iteration, 
the peak index obtained by 2D FFT method is recorded as ${l_1}$, as shown in Section \ref{sec3a}.
Phase compensation is performed for each element
of the channel information matrix ${\bf{Y}}$
according to the peak index ${l_1}$ in the first iteration.
Similiar to (\ref{eq25}), the $n$-th row of ${\bf{Y}}$ after phase compensation
is recorded as ${{\textbf{k}^\prime_{vn}} }$.
The DFT of ${{\textbf{k}^\prime_{vn}} }$ is calculated as follows.
\begin{equation}
	\begin{aligned}
		 {\bf{r}}\left( {{{\bar l}_d}} \right) & = \frac{1}{{{M_p} + {M_d}}} \times \\
		& \sum\limits_{m = 0}^{{{M_p} + {M_d}} - 1} {{e^{j2\pi m{T_{sym}}\left( {{f_d} - {f_{d1}}} \right)}}{e^{ - j2\pi m\frac{{{{\bar l}_d}}}{{{{\left( {{M_p} + {M_d}} \right)}^2}}}}}} \\
		&= \frac{1}{{{M_p} + {M_d}}}\sum\limits_{m = 0}^{{{M_p} + {M_d}} - 1} {{e^{j2\pi m\left( {{T_{sym}}{f_{d2}} - \frac{{{{\bar l}_d}}}{{{{\left( {{M_p} + {M_d}} \right)}^2}}}} \right)}}},
	\end{aligned}
\end{equation}
where ${\bar l}_d = 0,1,...,{M_p} + {M_d} - 1$,
${\bf{r}}\left( {{{\bar l}_d}} \right)$ is the ${\bar l}_d$-th element 
of the DFT of ${{\textbf{k}^\prime_{vn}} }$. 
Searching the peak of ${\bf{r}}\left( {{{\bar l}_d}} \right)$,
the peak index ${l_2}$ is obtained.
The estimated velocity with two iterations is
\begin{equation}
	V = \frac{c}{{2{f_c}{T_{sym}}}}  \left( {\frac{{{l_1} - 0.5 + \frac{{{l_2}}}{{{M_p} + {M_d}}}}}{{{M_p} + {M_d}}}} \right).
\end{equation}

Compared with the velocity estimation of 2D FFT method,
the accuracy of velocity estimation with $X$ iterations is
enhanced, which is as follows.

\begin{equation}
	\Delta V = \frac{c}{{2{f_c}{T_{sym}}{{\left( {{M_p} + {M_d}} \right)}^X}}}.
\end{equation}

\section{Long Distance Sensing}

In the scenario of long distance sensing, 
the CC algorithm is applied to
overcome the limitation of CP length on the sensing distance.
The concept of VCP is firstly proposed in \cite{14,17} to realize long distance sensing.
\cite{13} proposed another scheme for long distance sensing, 
where the CC algorithm and maximum likelihood (ML) algorithm are 
combined to construct the channel information matrix.
Then, the CC algorithm is used to estimate the distance 
and the 2D FFT method is used to estimate the velocity.
By increasing the number of samplings, 
the CC algorithm improves the accuracy of distance estimation.
However, the improvement of the accuracy of velocity estimation is not significant.
In order to improve the accuracy of velocity estimation,
we have proposed an iterative ISAC signal processing method in
Section \ref{sec3b}.
In the scenario of long distance sensing,
the ISAC signal combining pilot signal and phase coded data signal is applied,
which is processed by the iterative CC algorithm. 
In the scenario of long distance sensing, 
the SNR of echo signal is low, so that 
the phase coding improves the anti-noise performance
of ISAC signal.

\subsection{The CC Method with VCP and ML}

The transmitted ISAC signal $s\left( t \right)$ is sampled with the sampling interval ${T_b}$,
where ${T_b} = \frac{{{T_{sym}}}}{{N'}}$ and $N'$ is the number of samplings within an OFDM symbol. 
After removing the CP,
the sampling of transmitted signal is obtained.
The sampling points are divided into $\tilde M$ groups.
The number of sampling points in each group is $\tilde N$.
The received signal is sampled with the same sampling interval ${T_b}$.
Similarly, the sampling points are divided into $\tilde M$ groups.
And the number of sampling points in each group is $\tilde N$ \cite{14}.
The VCP is added, whose length is $\tilde Q$ \cite{14}.
VCP is the last $\tilde Q$ sampling points in a group, 
which are added to the first $\tilde Q$ sampling points of this group.
The $i$-th sampling point in the $\tilde m$-th group is shown as follows.

\begin{equation}\label{srrela}
	{r_{\tilde m}}\left[ i \right] = h{s_{\tilde m}}\left[ {i - \varepsilon } \right]{e^{j2\pi \upsilon \left( {\tilde m\tilde N + i} \right)}} + {\omega _{\tilde m}}\left[ i \right] + {I_{\tilde m}}\left[ i \right],
\end{equation}
where $\varepsilon  = \left\lfloor {\frac{\tau }{{{T_b}}}} \right\rfloor$, 
${I_{\tilde m}}\left[ i \right]$ is inter block interference due to VCP and follows Gaussian distribution \cite{14}, 
as shown in (\ref{eq36}).
Due to the addition of VCP, 
${\omega _{\tilde m}}\left[ i \right]$ follows Gaussian distribution \cite{14}, 
as shown in (\ref{eq35}).
\begin{equation}\label{eq36}
	{I_{\tilde m}}\left[ i \right] \sim N\left( {0,{\sigma ^2}} \right).
\end{equation}
\begin{equation}\label{eq35}
	{\omega _{\tilde m}}\left[ i \right] \sim \left\{ \begin{array}{l}
		N\left( {0,2{\sigma ^2}} \right),0 \le i < \tilde Q\\
		N\left( {0,{\sigma ^2}} \right),\tilde Q \le i < \tilde N
	\end{array} \right.,
\end{equation}

The correlation function $\rho \left( {p} \right)$ is constructed 
as follows
by calculating the correlation between received and transmitted signals \cite{13}.
\begin{equation}\label{huxiangguan}
	\rho \left( p \right) = \sum\limits_{i = 0}^{\tilde N - 1} {{r_{\tilde m}}\left[ i \right]s_{\tilde m}^*\left[ {i - p} \right]}.
\end{equation}

$\rho \left( p \right)$ is maximun when $p = \varepsilon$.
Searching the peak index $p = \varepsilon$, 
the delay $\tau$ is derived
by solving the equation $\varepsilon = \frac{\tau }{{{T_b}}}$. 
Then, the distance of target is estimated.

${{s_{\tilde m}}\left[ i \right]}$ is shifted by $\varepsilon$ samplings, 
yielding the shifted sequence ${{s_{\tilde m}}\left[ {i - \varepsilon } \right]}$.
The correlation function $\eta \left( {p,\tilde m} \right)$
of ${{s_{\tilde m}}\left[ {i - \varepsilon } \right]}$ and the original sequence ${{s_{\tilde m}}\left[ i \right]}$ is constructed as follows.
\begin{equation}\label{longcorr}
	\eta \left( {p,\tilde m} \right) = \sum\limits_{i = 0}^{\tilde N - 1} {{s_{\tilde m}}\left[ {i - \varepsilon } \right]s_{\tilde m}^*\left[ {i - p} \right]}.
\end{equation}

When calculating the correlation function $\eta \left( {p,\tilde m} \right)$, 
the received signal is coherently accumulated, 
increasing the SNR of the received echo signal by $10\log ( {\tilde N} )$ times, 
which improves the anti-noise ability of the ISAC signal \cite{18}.
According to the relation 
between ${r_{\tilde m}}\left[ i \right]$ and ${{s_{\tilde m}}\left[ i \right]}$ 
in (\refeq{srrela}),
the ML of $\varepsilon$ and $\upsilon$ is 
\begin{equation}\label{ML}
	\begin{aligned}
		& {\arg _{ML}}\left( {\varepsilon ,\upsilon } \right)  = \sum\limits_{\tilde m = 0}^{\tilde M - 1} {\sum\limits_{i = 0}^{\tilde N - 1} {{{\left| {\rho \left( {p} \right) - h{e^{j2\pi \upsilon \tilde m\tilde N}}\eta \left( {p,\tilde m} \right)} \right|}^2}} }\\
		&  = \underbrace {\sum\limits_{\tilde m = 0}^{\tilde M - 1} {\sum\limits_{i = 0}^{\tilde N - 1} {{{\left| {\rho \left( {p} \right)} \right|}^2}} } }_{\left( {\rm{a}} \right)} + \underbrace {\sum\limits_{\tilde m = 0}^{\tilde M - 1} {\sum\limits_{i = 0}^{\tilde N - 1} {{{\left| {h\eta \left( {p,\tilde m} \right)} \right|}^2}} } }_{\left( {\rm{b}} \right)}\\
		& - 2{\rm{Re}}\left( {\sum\limits_{\tilde m = 0}^{\tilde M - 1} {\sum\limits_{i = 0}^{\tilde N - 1} {\rho \left( {p} \right){{\left( {h{e^{j2\pi \upsilon \tilde m\tilde N}}\eta \left( {p,\tilde m} \right)} \right)}^*}} } } \right).
	\end{aligned}
\end{equation}

The terms (a) and (b) in (\refeq{ML}) 
have nothing to do with $\varepsilon$ and $\upsilon$,
which do not affect the result of ${\arg _{ML}}\left( {\varepsilon ,\upsilon } \right)$. 
When ${\arg _{ML}}\left( {\varepsilon ,\upsilon } \right)$ is minimum, 
(\refeq{MLtoFFT}) needs to take the maximum value. 
Therefore, 
the ML estimation of velocity is transformed into the maximization of $\mu \left( {p,\tilde m} \right)$ \cite{13}.
\begin{equation}\label{MLtoFFT}
	\mu \left( {p,\tilde m} \right) = \sum\limits_{\tilde m = 0}^{\tilde M - 1} {\sum\limits_{i = 0}^{\tilde N - 1} {\left| {\rho \left( {p} \right){\eta ^*}\left( {\varepsilon ,\tilde m} \right){e^{ - j2\pi \upsilon \tilde m\tilde N}}} \right|} }.
\end{equation}

When $p = \varepsilon  \buildrel \Delta \over = {p_0}$, 
${\eta ^{\rm{*}}}\left( {\varepsilon ,\tilde m} \right)$ 
is maximum, which is denoted by ${\eta _0}$. 
The ${p_0}$ is also the peak index of the correlation function $\rho \left( {p} \right)$ 
in (\refeq{huxiangguan}).
Similar to \cite{13}, 
letting $\upsilon =\frac{l}{{\tilde M\tilde N}},l = 0,1,...,\tilde M - 1$, (\refeq{MLtoFFT}) 
is transformed into the following form.
\begin{equation}\label{longvel}
	\mu \left( {{p_0},\tilde m} \right) = {\eta _0}\left| {\sum\limits_{\tilde m = 0}^{\tilde M - 1} {\rho \left( {p_0} \right)\exp \left( {{\rm{ - }}j2\pi \frac{{\tilde ml}}{{\tilde M}}} \right)} } \right|.
\end{equation}

The searching of the maximum value of $\mu \left( {{p_0},\tilde m} \right)$ 
respective to $l$ is regarded as the searching the 
peak index of the DFT of ${\rho \left( {{p_0}} \right)}$. 
The solution of maximization of $\mu \left( {{p_0},\tilde m} \right)$ is denoted by $l = l_0$. 
When $\mu \left( {{p_0},\tilde m} \right)$ is maximum, we have (\ref{eq42}) 
with the expansion of (\ref{longvel}),
and we have (\ref{eq43}).
\begin{equation}\label{eq42}
	\exp \left( {j2\pi \upsilon \tilde m\tilde N} \right)  \exp \left( { - j2\pi \frac{{\tilde m{l_0}}}{{\tilde M}}} \right) = 1.
\end{equation}

\begin{equation}\label{eq43}
	\varepsilon  = {p_0}.
\end{equation}

Solving the equation (\ref{eq42}), the $\upsilon$ is obtained as follows.

\begin{equation}
	\upsilon  = \frac{{{l_0}}}{{\tilde M\tilde N}}.
	\label{eq44}
\end{equation}
According to (\ref{eq43}) and (\ref{eq44}), the distance and velocity of target are calculated, respectively.
\begin{equation}
	R = \frac{{c{T_b}}}{2}{p_0},
\end{equation}
\begin{equation}
	V = \frac{c}{{2{f_c}{T_b}\tilde M\tilde N}}{l_0}.
\end{equation}

Then, we have the accuracy of distance and velocity estimation as follows.
\begin{equation}\label{eq47}
	\Delta R = \frac{{c{T_b}}}{2} = \frac{{c{T_{sym}}}}{{2N'}} = \frac{c}{{2B}}  \frac{N}{{N'}},
\end{equation}
\begin{equation}\label{eq48}
	\Delta V = \frac{c}{{2{f_c}\tilde N\tilde M{T_b}}} = \frac{c}{{2{f_c}\tilde M{T_{sym}}}}  \frac{{N'}}{{\tilde N}}.
\end{equation}

\subsection{The Iterative CC Method}

According to the algorithm mentioned in Section \ref{sec3b},
the accuracy of distance estimation 
is improved by increasing the number of sampling points in each OFDM symbol. 
However, in (\ref{eq48}), 
the product of $\tilde M$ and $\tilde N$ is $N'$,
so that we have $\Delta V = \frac{c}{{2{f_c}{T_{sym}}}}$,
which implies that 
reducing the sampling interval does not improve the accuracy of velocity estimation.
In this subsection,
an iterative CC method is proposed to
improve the accuracy of velocity estimation.
In the first iteration,
the velocity is estimated by the CC method.
Then, the channel information matrix undergoes phase compensation using the estimated velocity,
narrowing the interval of the velocity estimation.
The above operations are executed iteratively
to obtain a satisfactory accuracy of velocity estimation.

With $p = p_0$ and substituting (\refeq{huxiangguan}) into
(\refeq{longvel}), we have
\begin{equation}
	\begin{aligned}
		\mu \left( {{p_0},\tilde m} \right) = \sum\limits_{\tilde m = 0}^{\tilde M - 1} {{\eta _0}\exp \left( {j2\pi \upsilon \tilde m\tilde N} \right)\exp \left( {{\rm{ - }}j2\pi \frac{{\tilde ml}}{{\tilde M}}} \right)} .
	\end{aligned}
\end{equation}

The vector for velocity estimation in the first iteration is 
\begin{equation}
	\begin{aligned}
		{\bf{a}} & = \left[ {{A_0},...,{A_{\tilde m}}{e^{j2\pi \upsilon \tilde m\tilde N}},...,{A_{\tilde M - 1}}{e^{j2\pi \upsilon \left( {\tilde M - 1} \right)\tilde N}}} \right], \\
		\upsilon &= {\upsilon _0} + {\upsilon _1} + ... + {\upsilon _{X - 1}},
	\end{aligned}
\end{equation}
where ${\upsilon _0},{\upsilon _1},...,{\upsilon _{X - 1}}$ 
are the estimated values of $\upsilon$ in each iteration, 
and ${A_{\tilde m}}$ is
\begin{equation}
	{A_{\tilde m}} = h\sum\limits_{i = 0}^{\tilde N - 1} {{s_{\tilde m}}\left[ {i - \varepsilon } \right]s_{\tilde m}^*\left[ {i - p} \right]}, \tilde m = 0, 1, ..., \tilde M - 1.
\end{equation}

According to Section \ref{sec3b},
With $l = {l_0}$, 
the $l$-th element of the DFT of ${\textbf{a}}$ is maximum.
Hence, $l = {l_0}$ is the peak index in the first iteration.
The $\upsilon _0$ in the first iteration is 
\begin{equation}\label{eq52}
	{\upsilon _0} = \frac{{{l_0} - \frac{1}{2}}}{{\tilde M\tilde N}}.
\end{equation}

With (\ref{eq52}), 
$\textbf{a}$ undergoes phase compensation, yielding $\textbf{a}'$.
\begin{equation}
	{\bf{a'}} = {\bf{a}} \odot {\left[ {1,{e^{ - j2\pi {\upsilon _0}\tilde N}}, \cdots ,{e^{ - j2\pi {\upsilon _0}\left( {\tilde M - 1} \right)\tilde N}}} \right]}.
\end{equation}

Then, the second iteration starts, 
the DFT of $\textbf{a}'$ is calculated.
By searching the maximun value of the DFT of $\textbf{a}'$, 
the peak index $l=l_1$ is obtained.
With $l=l_1$, the ${\upsilon _1}$ is estimated in the second iteration.

\begin{equation}
	{\upsilon _1} = \frac{{{l_1} - \frac{1}{2}}}{{{{\tilde M}^2}\tilde N}}.
\end{equation}

With two iterations, the estimated velocity is expressed as
\begin{equation}
	V = \frac{c}{{2{f_c}{T_b}}}  \frac{{{l_0} - \frac{1}{2} + \frac{{{l_1}}}{{\tilde M}}}}{{\tilde M\tilde N}}.
\end{equation}

With $X$ iterations, the velocity estimation is recorded as $V$ 
and the accuracy of velocity estimation is $\Delta V$, 
which is improved compoared with 
CC method.
\begin{equation}
	V = \frac{c}{{2{f_c}\tilde N{T_b}}} \cdot \left( {\sum\limits_{\alpha  = 0}^{X - 2} {\frac{{{l_\alpha } - 0.5}}{{{{\tilde M}^{\alpha  + 1}}}}}  + \frac{{{l_{X - 1}}}}{{{{\tilde M}^X}}}} \right),
\end{equation}
\begin{equation}\label{eq57}
	\Delta V = \frac{c}{{2{f_c}\tilde N{T_b}{{\tilde M}^X}}}.
\end{equation}

When the channel information matrix is obtained, 
the 2D FFT method calculates $M$ times of $N$-point IDFT and $N$ times of $M$-point 
DFT to estimate the distance and velocity of the target, respectively. 
Therefore, the computational complexity of 2D FFT algorithm 
is ${\rm O}\left( {{M^{\rm{2}}}{\rm{ + }}{N^{\rm{2}}}} \right)$. 
The CC algorithm divides the samplings of transmitted signal and 
received signal into ${\tilde M}$ groups, 
with ${\tilde N}$ sampling points in each group. 
The correlation function of a group of samplings of the transmitted signal and 
the received signal is calculated, 
and then the $M$-point DFT is calculated. 
Therefore, the computational complexity of CC algorithm 
is ${\rm O}\left( {{{\tilde M}^{\rm{2}}} + {{\tilde N}^{\rm{2}}}} \right)$.
Supposing that the iterative 2D FFT method has $X$ iterations, 
there are $M$ times of 
$N$-point IDFT and $N$ times of $M$-point DFT in each iteration. 
Therefore, the computational complexity of the iterative 2D FFT method
is ${\rm{O}}\left( {X * \left( {{M^{\rm{2}}}{\rm{ + }}{N^{\rm{2}}}} \right)} \right)$. 
When sensing accuracy of the method in \cite{15} is the same to 
iterative 2D FFT method, 
the computational complexity of the method in \cite{15} 
is ${\rm{O}}\left( {{{\left( {{M^{\rm{2}}}{\rm{ + }}{N^{\rm{2}}}} \right)}^X}} \right)$,
which is higher than the iterative 2D FFT method proposed in this paper.

\section{Simulation Results and Analysis}

In this section,
the performance of ISAC signal
and the performance of
iterative signal processing algorithms are verified.
The main parameters in Subsections B and C are shown in Table \ref{JSC signal parameters}.

\begin{table}[htb]
	\renewcommand{\arraystretch}{1.5}
	\begin{center}
		\caption{Parameters of ISAC signal }
		\label{JSC signal parameters}
		\setlength{\tabcolsep}{8mm}{
			\begin{tabular}{|c|c|}
				\hline   \textbf{Parameters} & \textbf{Values} \\
				\hline   Interval of subcarrier & 15 kHz \\
				\hline   Carrier frequency & 24 GHz \\
				\hline   Number of subcarriers & 256 \\
				\hline   Numbers of symbols & 14 \\
				\hline   Sampling interval & 0.13 $\mu s$ \\
				\hline
			\end{tabular}
		}
	\end{center}
\end{table}

\subsection{Ambiguity Function}

\begin{figure}[htbp]
	\centering
	\includegraphics[width=0.49\textwidth]{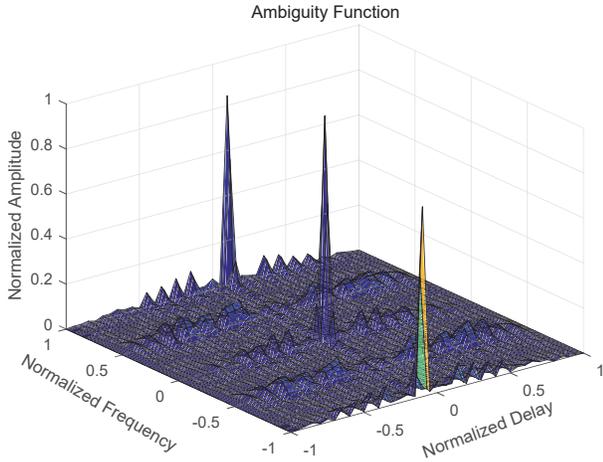}
	\caption{AF of ISAC signal with phase coding using Golay sequence.}
	\label{AFPC}
\end{figure}

\begin{figure}[htbp]
	\centering
	\includegraphics[width=0.49\textwidth]{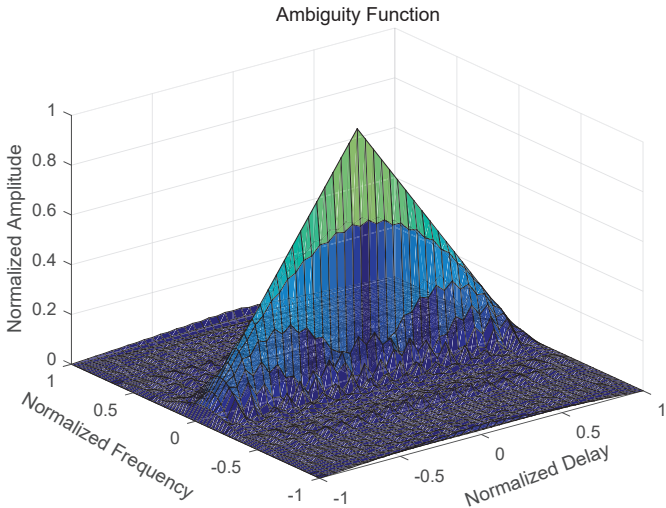}
	\caption{AF of ISAC signal without phase coding.}
	\label{AFBPSK}
\end{figure}

Fig. \ref{AFPC}
illustrates the ambiguity function (AF) of the ISAC signal combining pilot and data symbols 
using 32 bit Golay sequence in phase coding.
Fig. \ref{AFBPSK} shows
the AF of ISAC signal without phase coding.
The number of subcarriers of ISAC signal is 32.
The subcarrier interval is set to 15 kHz.
The number of OFDM symbols is 10.
Comparing Fig. \ref{AFPC} and Fig. \ref{AFBPSK},
it is discovered that the ISAC signal encoded with Golay sequence has a pushpin AF,
while the AF of the ISAC signal without phase coding has a triangular envelope.
The sidelobe attenuation of the AF of
the ISAC signal without phase coding is
slower than the AF of the ISAC signal with phase coding.
Therefore,
phase coding effectively improves the sensing performance of ISAC signal.

\subsection{Short range sensing}

\begin{figure}
	\centering
	\includegraphics[width=0.49\textwidth]{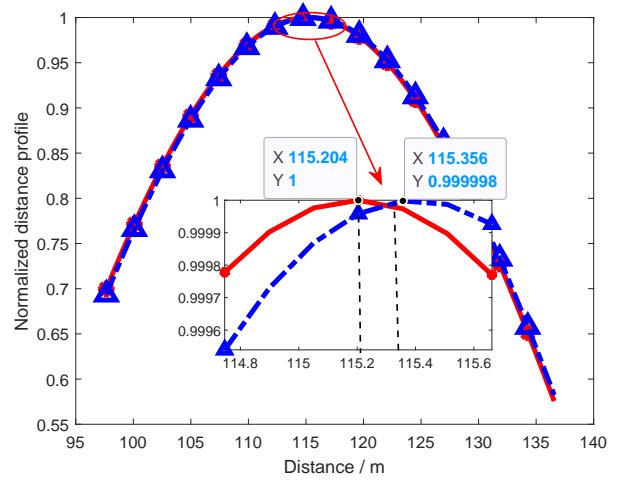}
	\caption{Distance estimation performance of proposed ISAC signal processing method in short distance sensing.}
	\label{Distance short}
\end{figure}

\begin{figure}
	\centering
	\includegraphics[width=0.49\textwidth]{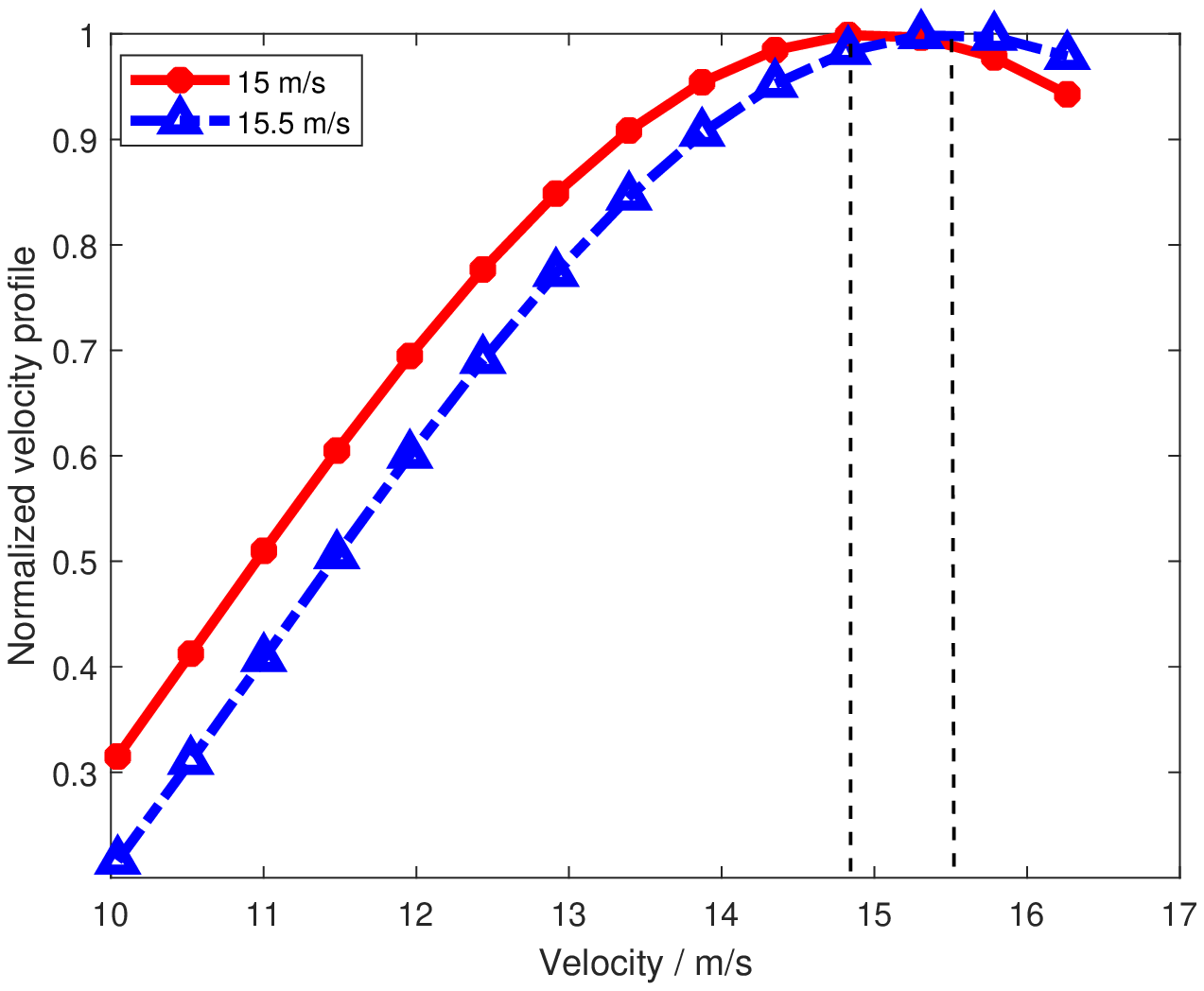}
	\caption{Velocity estimation performance of proposed ISAC signal processing method in short distance sensing.}
	\label{Velocity short}
\end{figure}

Two targets are detected in the scenario of short range sensing
to verify the performance
improvement of iterative 2D FFT algorithm in
Section \uppercase\expandafter{\romannumeral3}.
The distance of the two targets is 115.2 m and 115.4 m, respectively,
and the velocity of the two targets is 15 m/s and 15.5 m/s, respectively.
Using the iterative 2D FFT algorithm proposed in Section \uppercase\expandafter{\romannumeral3},
the sensing results are shown in Fig. \ref{Distance short} and Fig. \ref{Velocity short}.
It is revealed that the error of distance and velocity estimation is 
reduced to 0.037 \% and 1.2 \%, respectively.

\begin{figure}
	\centering
	\includegraphics[width=0.49\textwidth]{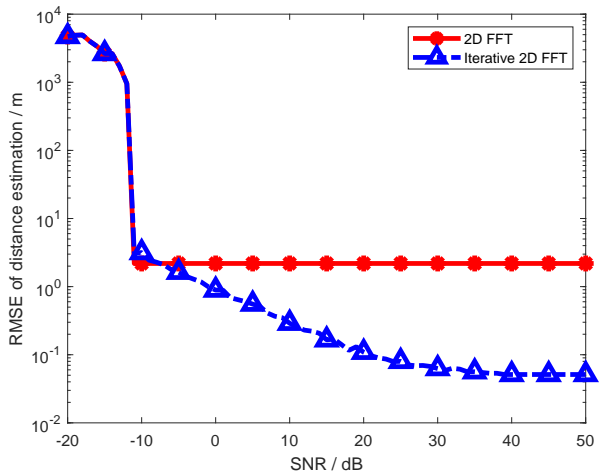}
	\caption{Distance estimation performance of 2D FFT and iterative 2D FFT methods.}
	\label{suanfaduibiR}
\end{figure}

\begin{figure}
	\centering
	\includegraphics[width=0.49\textwidth]{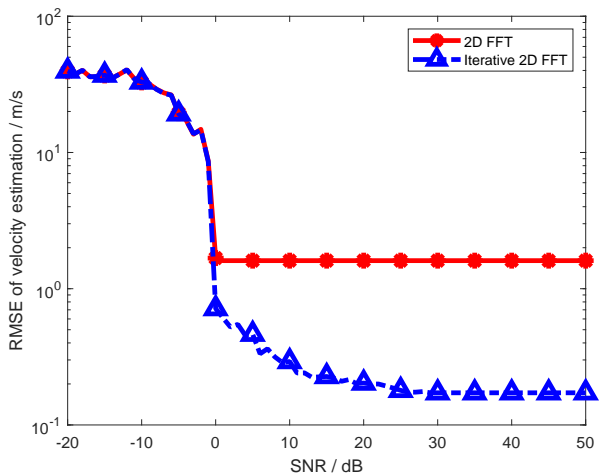}
	\caption{Velocity estimation performance of 2D FFT and iterative 2D FFT methods.}
	\label{suanfaduibiV}
\end{figure}

In the iterative 2D FFT algorithm,
the number of iterations is set as 2.
The Root Mean Square Errors (RMSEs) of
the estimation of distance and velocity are shown in
Fig. \ref{suanfaduibiR} and Fig. \ref{suanfaduibiV}, respectively.
With low SNR,
the sensing performance of the two algorithms is similar.
With the increase of SNR,
the RMSE of iterative 2D FFT algorithm is lower than that of 2D FFT algorithm
and the gap between them is increasing.
Simulation results verified that the proposed iterative 2D FFT algorithm
has higher accuracy and effectively overcomes the quantization error
of 2D FFT algorithm.
It is revealed from the Fig. \ref{suanfaduibiR} and Fig. \ref{suanfaduibiV} that 
RMSE first decreases with the increase of SNR, and finally converges to a constant value.
This is due to the fact that the estimated values of distance and velocity are 
related to the peak indexes at each iteration. 
However, these peak indexes take integer, 
so that there is an error between the estimated value and the real value, 
which is called quantization error. 
With the increase of the number of iterations, 
the quantization error decreases. 
With the increase of SNR, the estimated values of 
distance and velocity will become more and more accurate and gradually approach the real value. 
Therefore, RMSE will decrease with the increase of SNR. 
When the SNR increases to a threshold, 
the impact of noise on the estimated 
value is small and can be ignored. 
Meanwhile, 
RMSE no longer decreases with the increase of SNR since it is mainly 
resulted from quantization error.


\subsection{Long range sensing}

\begin{figure}
	\centering
	\includegraphics[width=0.49\textwidth]{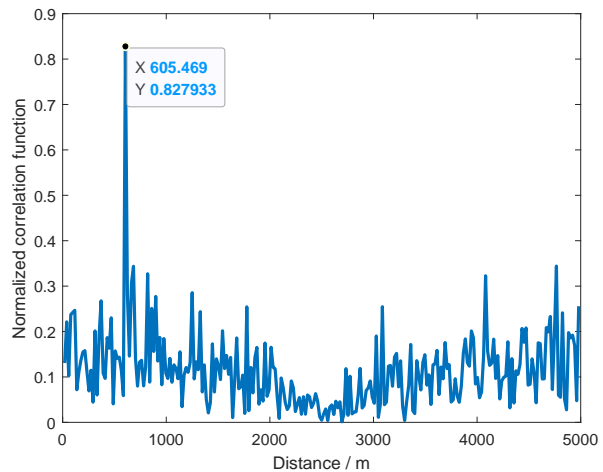}
	\caption{Distance estimation in the scenario of long distance sensing.}
	\label{Distance long}
\end{figure}

\begin{figure}
	\centering
	\includegraphics[width=0.49\textwidth]{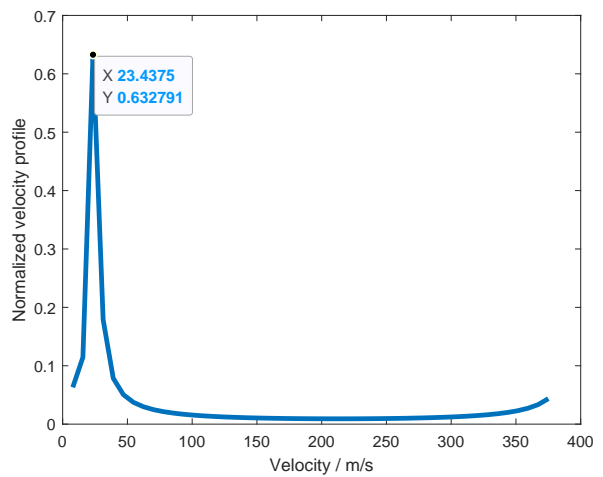}
	\caption{Velocity estimation in the scenario of long distance sensing.}
	\label{Velocity long}
\end{figure}

The performance of ISAC signal processing method in the scenario of long distance sensing
is verified in this subsection.
According to the simulation parameters in Table \ref{JSC signal parameters},
the target with velocity of 25 m/s and distance of 600 m is detected.
The sensing results of distance and velocity using
the iterative CC algorithm are shown in Fig. \ref{Distance long} and
Fig. \ref{Velocity long}.
The estimated distance and velocity of target are 605.5 m and 23.44 m/s, respectively,
with errors of 0.84 \% and 6.24 \%, respectively.
It is revealed that the iterative CC algorithm
correctly estimates the sensing information.


\begin{figure}
	\centering
	\includegraphics[width=0.49\textwidth]{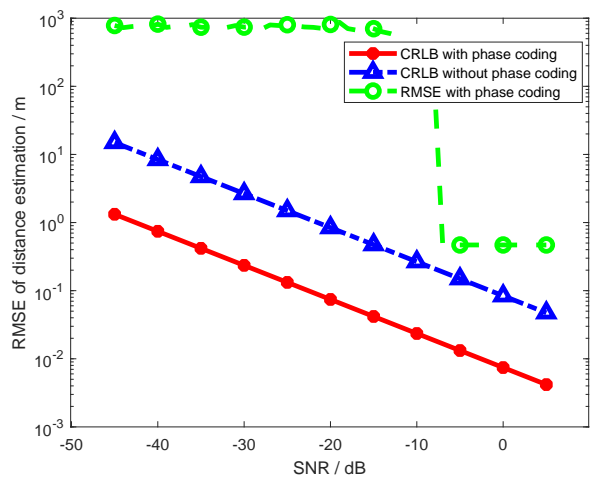}
	\caption{Distance estimation performance of proposed ISAC signal.}
	\label{Distancermse}
\end{figure}

\begin{figure}
	\centering
	\includegraphics[width=0.49\textwidth]{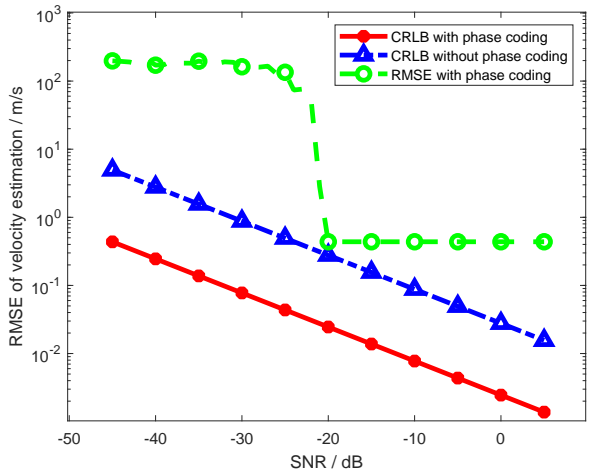}
	\caption{Velocity estimation performance of proposed ISAC signal.}
	\label{Velocityrmse}
\end{figure}

When 32-bit Golay sequence is applied to phase encode the ISAC signal,
the RMSE and CRLB of distance and velocity estimation are shown in
Fig. \ref{Distancermse} and Fig. \ref{Velocityrmse}, respectively.
It is revealed that the ISAC signal with phase coding has better anti-noise performance,
yealding a lower CRLB compared with the ISAC signal without phase coding.


\begin{figure}
	\centering
	\includegraphics[width=0.49\textwidth]{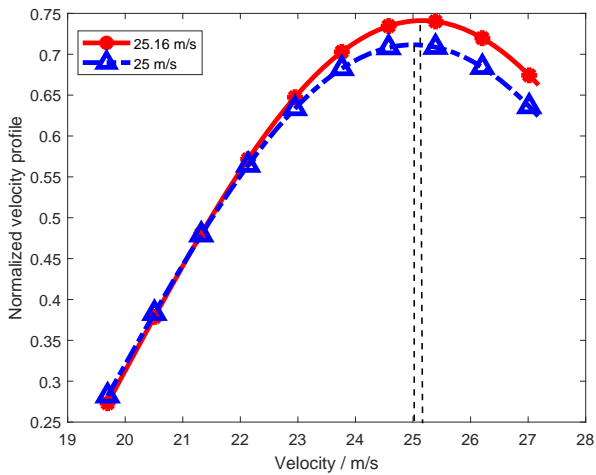}
	\caption{Velocity estimation performance in long distance sensing.}
	\label{VCPvel}
\end{figure}

With the parameters in
Table \ref{JSC signal parameters},
the velocity accuracy is 7.8125 m/s without iteration according to (\ref{eq48}),
and the velocity accuracy of the iterative CC algorithm is improved to 
0.16 m/s with two iterations
according to (\ref{eq57}).
As shown in Fig. \ref{VCPvel},
the velocity of the two targets is 25 m/s
and 25.16 m/s, respectively.
The CC algorithm cannot distinguish the two targets,
while the two targets can still be distinguished with the iterative CC algorithm.

\section{Conclusion}

In this paper, in order to improve the anti-noise performance of ISAC signal,
phase coding is applied in ISAC signal design combining the pilot and data signals.
Then, the CRLB of the ISAC signal is analyzed.
The iterative ISAC signal processing algorithms with high accuracy and low-complexity are proposed.
Taking the short distance and long distance sensing scenarios into account,
the iterative ISAC signal processing methods are proposed based on 2D FFT
and CC algorithms, respectively.
This paper provides the iterative ISAC signal processing algorithms 
with high accuracy and low complexity,
which is promising to be applied in the energy efficient ISAC systems.

\bibliographystyle{IEEEtran}
\bibliography{iterative}

\ifCLASSOPTIONcaptionsoff
  \newpage
\fi

\vspace{-10 mm}

\end{document}